\documentclass[Physsubmission, Phys]{SciPost}

\newcommand{\PZ}{$\pi^{0}\text{}$}
\newcommand{\EToP}{$\eta/\pi^{0}\text{}$}
\newcommand{\pt}{$p_{\text{T}}\text{}$}
\newcommand{\RpA}{$R_{\text{pPb}}$}
\newcommand{\pPb}{p--Pb at $\sqrt{s_{\text{NN}}}~=~\text{8.16 TeV}$}
\newcommand{\pp}{pp at $\sqrt{s}~=~\text{8 TeV}$}

\hypersetup{
    colorlinks,
    linkcolor={red!50!black},
    citecolor={blue!50!black},
    urlcolor={blue!80!black}
}

\usepackage[bitstream-charter]{mathdesign}
\DeclareSymbolFont{usualmathcal}{OMS}{cmsy}{m}{n}
\DeclareSymbolFontAlphabet{\mathcal}{usualmathcal}

\begin{document}

\begin{center}{\Large \textbf{\boldmath
Measuring the light meson nuclear modification factor in p--Pb collisions over an unprecedented $p_{\text{T}}$ range with ALICE\\
}}\end{center}

\begin{center}
Joshua König\textsuperscript{1$\star$} for the ALICE collaboration
\end{center}

\begin{center}
{\bf 1} Goethe-Universität Frankfurt am Main
\\
* joshua.konig@cern.ch
\end{center}

\begin{center}
\today
\end{center}

\linenumbers

\definecolor{palegray}{gray}{0.95}
\begin{center}
\colorbox{palegray}{
  \begin{tabular}{rr}
  \begin{minipage}{0.1\textwidth}
    \includegraphics[width=22mm]{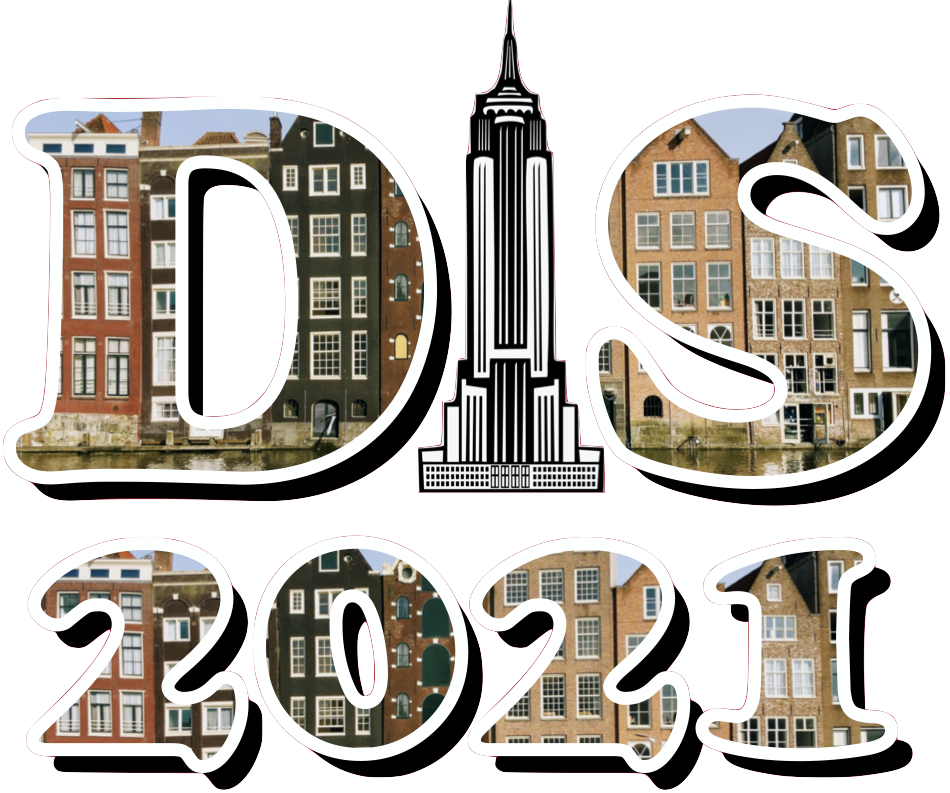}
  \end{minipage}
  &
  \begin{minipage}{0.75\textwidth}
    \begin{center}
    {\it Proceedings for the XXVIII International Workshop\\ on Deep-Inelastic Scattering and
Related Subjects,}\\
    {\it Stony Brook University, New York, USA, 12-16 April 2021} \\
    \doi{10.21468/SciPostPhysProc.?}\\
    \end{center}
  \end{minipage}
\end{tabular}
}
\end{center}

\section*{Abstract}
{\bf
Differential invariant cross sections of light neutral mesons in p--Pb collisions at $\sqrt{s_{\text{NN}}}$ = 8.16 TeV and in pp collisions at $\sqrt{s}$ = 8 TeV  have been measured up to very high transverse momentum (\pt). By combining independent reconstruction techniques available in ALICE using the EMCal and PHOS calorimeters as well as the central barrel tracking detectors, the combined spectra cover almost two orders of magnitude in \pt~for the \PZ~meson. The nuclear modification factor \RpA~has been measured for the \PZ~and $\eta$~mesons and is found to be consistent with NLO pQCD, CGC and energy loss calculations. Comparisons to the \RpA~of \PZ~measured in $\sqrt{s_{\text{NN}}}$ = 5.02 TeV hint at a stronger suppression at low \pt~with increasing collision energy.
}

\noindent\rule{\textwidth}{1pt}
\vspace{10pt}

\section{Introduction}
\label{sec:intro}
Ultrarelativistic collisions of protons and nuclei provide an ideal environment to study the influence of intial-state effects on particle production. In contrast to Pb-Pb collisions it is expected that in p--Pb collisions the energy density is not high enough to form a quark-gluon plasma (QGP). Modifications of the particle production in p--Pb collisions compared to pp collisions can therefore be attributed to a modification of the parton distribution functions (PDFs), describing the fractional momentum ($x$) of the partons in the nucleon.
Measurements of the PDFs show that in a nucleus, the PDFs of the partons (nPDFs) are modified compared to the PDF of a single proton or neutron: At small $x$, a significant depletion of the nPDFs is observed, commonly known as shadowing.
In addition to nuclear shadowing, effects of gluon saturation in the heavy nucleus can be described by the Color-Glass-Condensate (CGC) model \cite{cgcIntro}. Furthermore, parton energy loss in the cold nuclear matter can also play a role in the modification of particle production \cite{fcelIntro}.\\
By comparing particle production in pp collisions and p--Pb collisions at the same center of mass energy, the influence of the nuclear environment can be measured via the nuclear modification factor:
\begin{equation}
	R_{\text{pA}} = \frac{1}{A} \frac{d^2 \sigma_{Pb}}{dp_{T}dy} / \frac{d^2 \sigma_{pp}}{dp_{T}dy}
\end{equation}
where $A$ is the nuclear mass number, $d^2 \sigma_{Pb}/dp_{T}dy$ the measured cross section in p--Pb collisions and $d^2 \sigma_{pp}/dp_{T}dy$ the measured cross-section in pp collisions at the same center of mass energy. Any deviation from unity indicates a modification in particle production in p--Pb compared to pp.

\section{Detector setup and datasets}
%
%
%
The neutral meson measurements were performed using the dominant decay channel of the \PZ~and $\eta$ meson into two photons. The photons can be reconstructed with the ALICE detector system either via one of the electromagnetic calorimeters or with the photon-conversion method (PCM). The latter makes use of the conversion probability of photons of about 8.9\% before they reach the main tracking detector, the time projection chamber (TPC). These converted photons can be reconstructed by measuring the resulting $e^{\pm}$ tracks with the central tracking detectors ITS (inner tracking system) and TPC, providing an excellent energy resolution down to very low $p_{\text{T}}$. Furthermore, photons can be measured using the electromagnetic calorimeter (EMCal), which provides a large acceptance and can measure photons up to very high energies. Additionally, the photon spectrometer (PHOS) complements the EMCal, having a fine cell granularity and therefore providing a good energy resolution. 
A detailed description of the detector systems can be found in \cite{alice,upgrades}.\\
The datasets used for the analysis presented in this article are from p--Pb collisions at $\sqrt{s_{\text{NN}}}$~= 8.16 TeV recorded in 2016 during the LHC run 2 data taking and pp collisions at $\sqrt{s}$ = 8 TeV recorded in 2012 during the LHC run 1.
The data was recorded using the minimum-bias trigger which relies on a coincident signal in both V0 detectors. Additionally calorimeter triggers are used, which are based on a large energy deposit in the EMCal or PHOS in a small array of cells. Using these triggers, integrated luminosities of $\mathcal{L} =$~11~nb$^{-1}$ in p--Pb and $\mathcal{L} =$~657~nb$^{-1}$ in pp are obtained.


\section{Neutral Meson reconstruction}
The reconstruction of the neutral mesons from their decay photons is performed using invariant mass based methods and a purity based method.
\subsection{Invariant Mass based methods}
\label{invMass}
Using the energies ($E_{\text{1}}, E_{\text{2}}$) and positions of the measured photons, the invariant mass ($m_{\text{inv}}$) for each photon-pair in the event can be calculated using $ m_{\text{inv}} = \sqrt{2E_{\gamma 1}E_{\gamma 2}\cdot(1-cos(\Theta_{\text{1,2}}))}$, where $\Theta_{\text{1,2}}$ is the opening angle between the two photons.
Both photons can be measured by the same photon reconstruction technique (PCM, EMCal, PHOS) or a hybrid approach (PCM-EMCal) is used, where one photon is reconstructed with PCM and one is measured with the EMCal.
The background is estimated using a mixed-event technique. 
After background subtraction, the raw yield is obtained by integration of the $m_{\text{inv}}$ distribution around the meson mass which is estimated by a combined parametrization of a Gaussian and exponential function. 

\begin{figure}[t!]
	\centering
	\includegraphics[width=0.8\textwidth]{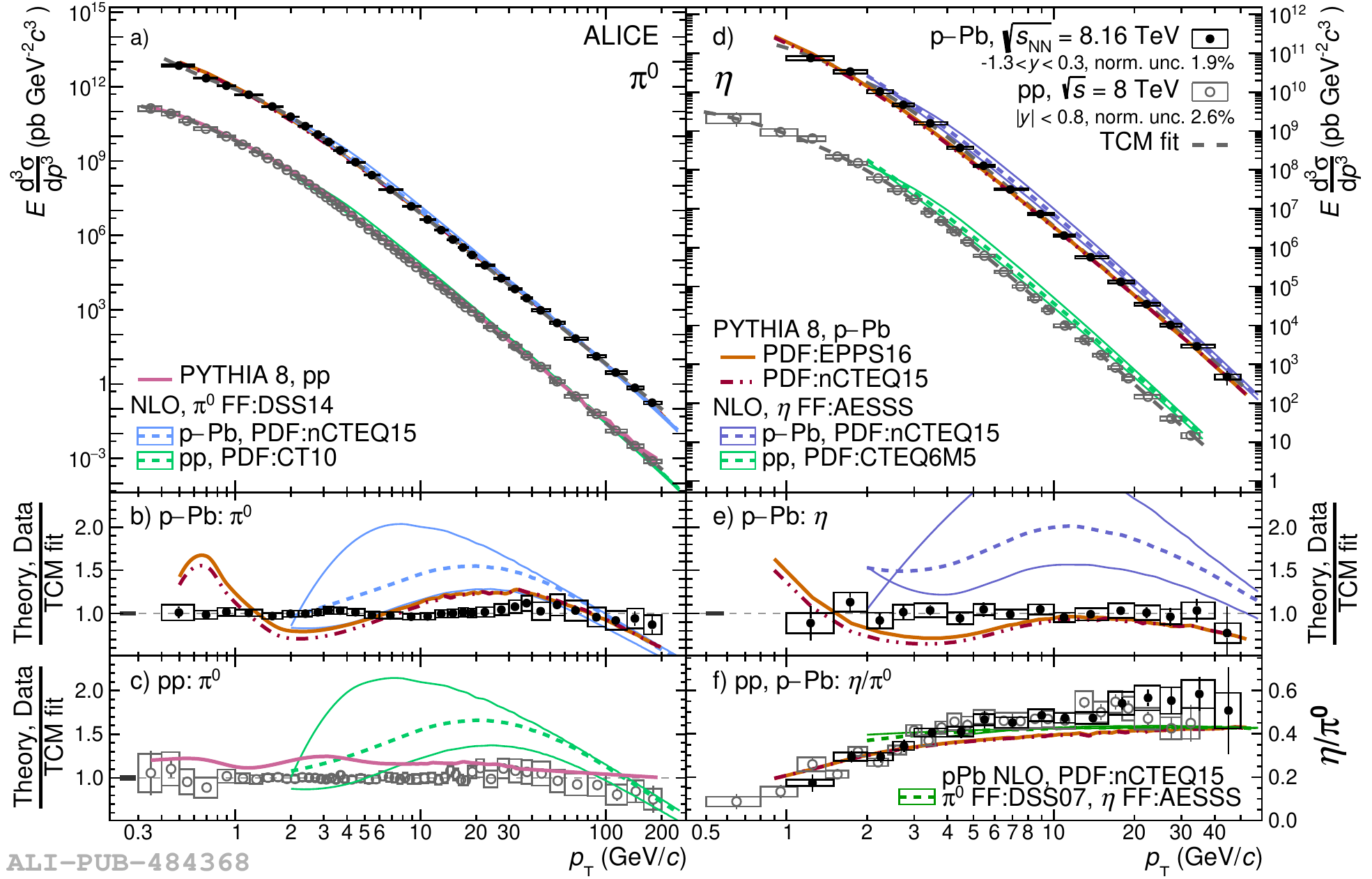}
	
	\caption{\PZ~(a) and $\eta$ (d) meson differential invariant cross section for pp at $\sqrt{s} = \text{8 TeV }$ and for p--Pb at $\sqrt{s_{\text{NN}}} = \text{8.16 TeV }$ together with PYTHIA~8 and NLO calculations. (b), (c) and (e) show ratios of the data and theory calculations to the two component model (TCM) parametrizations \cite{tcm} and (f) shows the \EToP~ratio for both pp and p--Pb collisions together with theory calculations. Figure from \cite{pPb8TeV}.}
	\label{fig:spectra}
\end{figure}
\subsection{Purity based methods}
\label{merged}
With rising \pt, the opening angle of the decay photons of a \PZ~meson decreases. Beyond \pt~$\approx$~16 GeV/$c$, the cell granularity of the EMCal does not allow separating the two photon showers anymore and as a result, only a single cluster containing both \PZ~decay photons is measured. To be able to reconstruct neutral pions with the EMCal up to very high \pt, the merged clusters have to be selected and the resulting raw yield has to be corrected for contamination. The merged \PZ~clusters typically have an elliptical shape compared to circle-shaped single photon clusters. The shape is quantified by $\sigma_{long}^{2}$ which can be interpreted as the long axis of the cluster ellipse. A cut of $\sigma_{long}^{2} > 0.27$ is used to select merged \PZ~candidates while rejecting a large fraction of single photon clusters. The purity of the selected cluster sample exceeds 80\% in all analyzed \pt~intervals from \pt~=~16 to 200 GeV/$c$.

\subsection{Corrections and Combination}
The raw \PZ~and $\eta$ meson yields are corrected for detector effects and contamination using the PYTHIA~8 Monte Carlo event generator together with GEANT3 detector simulations. In addition to acceptance, reconstruction efficiency and purity corrections, the \PZ~spectra are corrected for secondary decays from $K^{0}_{s}$, $K^{0}_{L}$ and $\Lambda$.\\
The different invariant \PZ~and $\eta$ spectra obtained with the reconstruction techniques described in section \ref{invMass} and \ref{merged} are combined using the BLUE \cite{blue} method which takes the statistical and systematic uncertainties into account.



\section{Results}
Fig. \ref{fig:spectra} shows the differential invariant \PZ~(a) and $\eta$ (d) meson cross sections in p--Pb collisions at \mbox{$\sqrt{s_{\text{NN}}}$ = 8.16 TeV} and in pp collisions at $\sqrt{s}$ = 8 TeV \cite{pPb8TeV}. In p--Pb the \PZ~($\eta$) spectrum covers 0.4~$ \leq p_{\text{T}} < $~200 GeV/$c$ (1.0 $ \leq p_{\text{T}} < $ 50 GeV/$c$) making it the  highest \pt-reach for identified particles and $\eta$ mesons to date. The pp \PZ~reference measurement was extended to \pt~=~200~GeV/$c$ to match the \pt-reach of the p--Pb spectrum for the calculation of the \RpA.
The spectra obtained in p--Pb collisions are compared to pQCD NLO calculations using nCTEQ15 \cite{nCTEQ} together with DSS14 \cite{dss} for the \PZ and nCTEQ15 together with AESSS \cite{aess} for the $\eta$. Furthermore, comparisons to PYTHIA 8 using nCTEQ15 are shown. While the NLO calculations overshoot the data especially for the $\eta$, PYTHIA 8 matches the absolute magnitude of the data better.
Using the \PZ~and $\eta$ meson spectra, the \EToP~ratio can be calculated and is shown in Fig. \ref{fig:spectra}(f) for \pPb~and for \pp. The two ratios are in very good agreement and a high \pt~ constant fit above 3 GeV/$c$ gives a value of 0.48 $\pm$ 0.01 for p--Pb and 0.47 $\pm$ 0.01 for pp. Both the NLO calculation as well as PYTHIA 8 predict a slightly lower \EToP~ratio.

\subsection{Nuclear modification factor}
\begin{figure}[t!]
	\centering
	\includegraphics[width=0.5\textwidth]{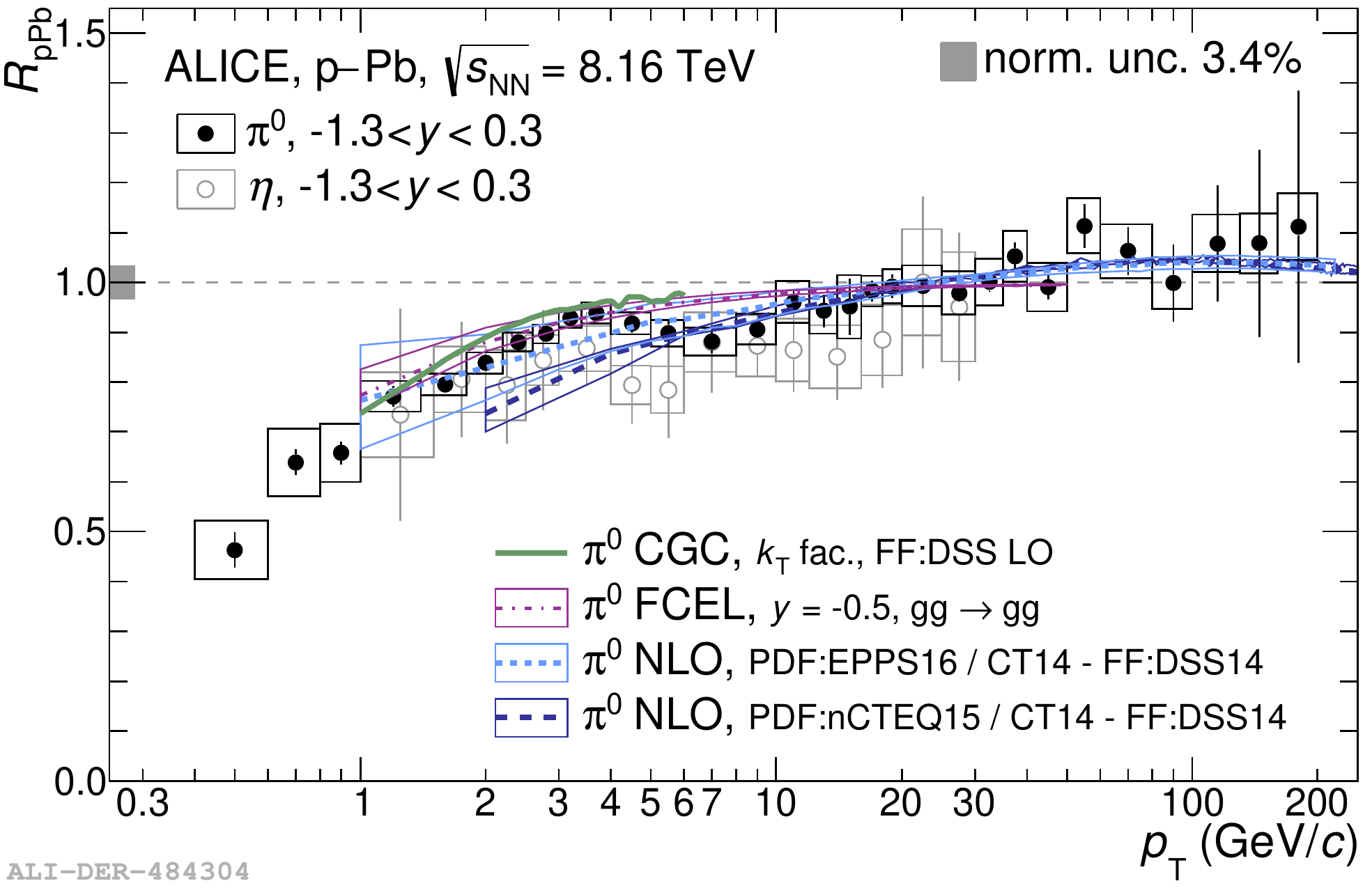}
	\caption{\RpA~for \PZ~and $\eta$ mesons for \pPb~as function of \pt~together with CGC and FCEL calculations and two NLO calculations using different nPDFs. Figure from \cite{pPb8TeV}.}
	\label{fig:Rpa}
\end{figure}
Fig. \ref{fig:Rpa} shows the measured \RpA~for \PZ~and $\eta$ mesons in \pPb~as function of \pt~\cite{pPb8TeV} together with pQCD NLO calculations using nPDFs EPPS16 \cite{epps} and nCTEQ15 as well as a CGC \cite{cgc} and an FCEL \cite{fcel} calculation. The pp reference spectrum at $\sqrt{s}$ = 8 TeV was corrected for the energy and rapidity shift to match the \pPb~measurement. The \PZ~measurement shows a strong suppression at low \pt~which is described by all calculations except the NLO calculation using the nCTEQ15 nPDF which predicts a lower \RpA~compared to the data. At \pt~$\approx$ 3 GeV/$c$ a Cronin-peak-like structure is visible, however it is not as pronounced as for charged hadrons \cite{alice5TeVhadrons}. Above \pt~=~10 GeV/$c$ the \RpA~of \PZ~is compatible with unity which is consistent with all theoretical predictions shown. The $\eta$ meson \RpA~is in agreement with the \PZ measurement within the uncertainties.

\subsection{Comparison to other measurements}
\begin{figure}[t!]
	\centering
	\includegraphics[width=0.45\textwidth]{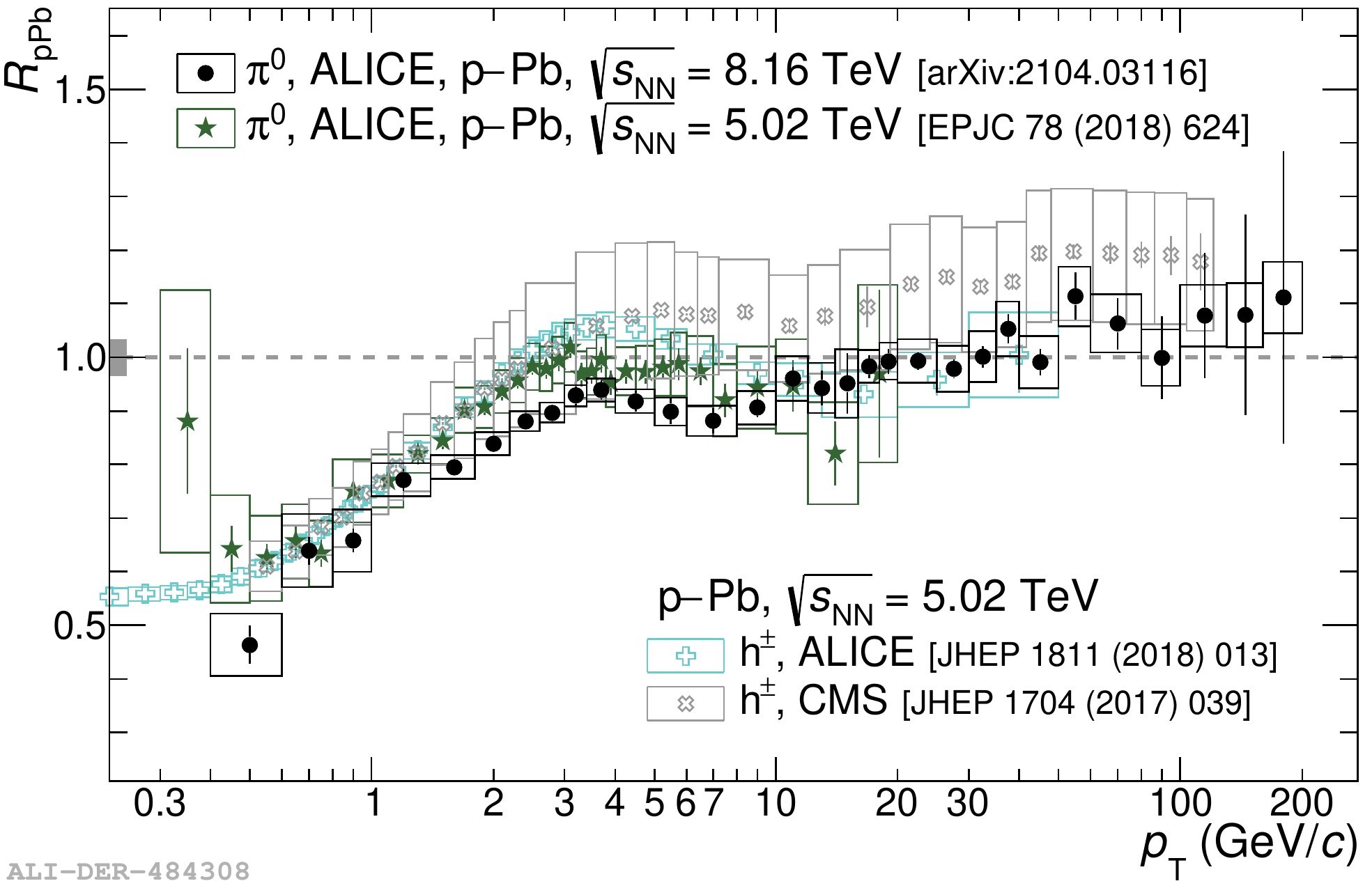}
	\includegraphics[width=0.45\textwidth]{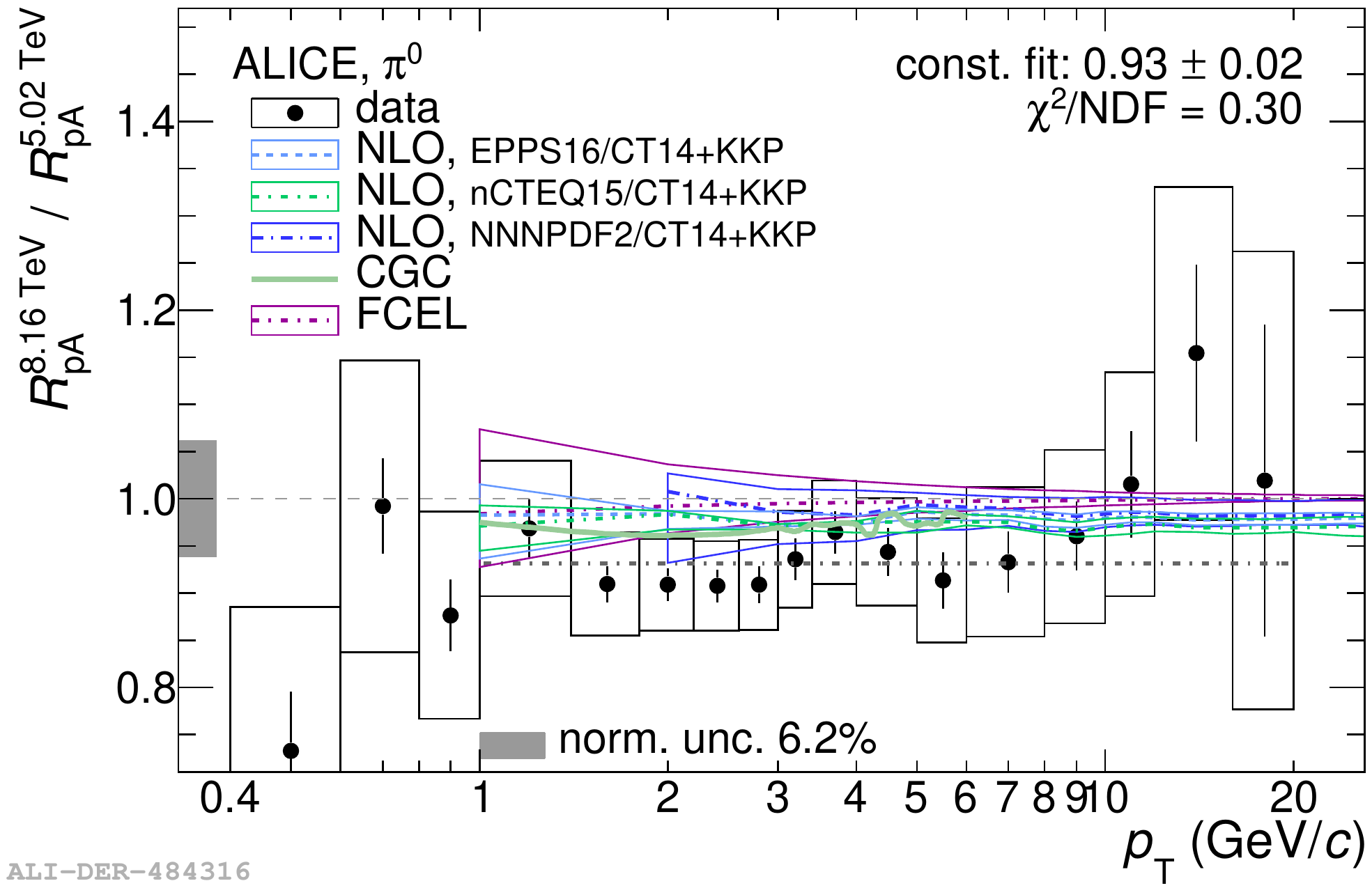}
	\caption{Left: \RpA~for neutral pions at $\sqrt{s_{\text{NN}}}$ = 8.16 TeV and at $\sqrt{s_{\text{NN}}}$ = 5.02 TeV as well as the \RpA~of charged hadrons measured with ALICE and CMS at $\sqrt{s_{\text{NN}}}$ = 5.02 TeV. Right: Ratio of the \RpA~in $\sqrt{s_{\text{NN}}}$ = 8.16 TeV to $\sqrt{s_{\text{NN}}}$ = 5.02 TeV together with theory calculations. Figures from \cite{pPb8TeV}.}
	\label{fig:Rpawith5}
\end{figure}

Fig. \ref{fig:Rpawith5} (a) shows \RpA~ for \PZ~at $\sqrt{s_{\text{NN}}} = \text{8.16 TeV }$ \cite{pPb8TeV} compared to \RpA~for \PZ and charged hadrons at $\sqrt{s_{\text{NN}}} = \text{5.02 TeV}$ \cite{alice5TeV, alice5TeVhadrons, cms}. The charged hadron measurement exhibits a larger Cronin-peak-like structure compared to both \PZ~measurements  which is attributed to a stronger Cronin effect for baryons. From \pt~=~10 GeV/$c$ onward, the measurements of ALICE are in agreement and are compatible with unity. The charged hadron measurement from CMS indicates a higher \RpA~than the \PZ~measurement at $\sqrt{s_{\text{NN}}} = \text{8.16 TeV}$, however the two measurements are still compatible within their respective uncertainties.
To study a possible energy dependence of the \RpA, the ratio of the \RpA~for \PZ~at $\sqrt{s_{\text{NN}}} = \text{8.16 TeV}$ and at $\sqrt{s_{\text{NN}}} = \text{5.02 TeV}$ is shown in Fig. \ref{fig:Rpawith5} (b) \cite{pPb8TeV}. A constant fit to the data gives 0.93 $\pm$ 0.02 but taking the normalization uncertainty of 6.2~\% into account, which mostly originates from an interpolated \PZ~reference measurement for the $\sqrt{s_{\text{NN}}} = \text{5.02 TeV}$ results, the two \RpA~measurements are compatible. The theory calculations predict a mild energy dependence with at most 2-3~\% difference between the two collision energies.



\section{Conclusion}
The \PZ~and $\eta$ meson cross sections for \pPb~have been measured up to very high transverse momentum. The \PZ~reference measurement for \pp~has been extended to match the \pt~reach of the p--Pb measurement. The \RpA~at $\sqrt{s_{\text{NN}}} = \text{8.16 TeV }$ exhibits a strong suppression at low \pt~and no suppression or enhancement at high \pt. It is compatible with most of the pQCD NLO, CGC and energy loss calculations. A comparison to the \RpA~at $\sqrt{s_{\text{NN}}} = $ 5.02 TeV hints at a larger suppression at low \pt~with rising collision energy however the data is still compatible within the given uncertainty.

\section*{Acknowledgements}
We thank W. Vogelsang, T. Lappi and H. Mäntysaari and  F. Arleo et. al for proving the theory calculations.
\\
The ALICE Collaboration would like to thank all its engineers and technicians for their invaluable contributions to the construction of the experiment and the CERN accelerator teams for the outstanding performance of the LHC complex. The ALICE Collaboration gratefully acknowledges the resources and support provided by all Grid centres and the Worldwide LHC Computing Grid (WLCG) collaboration.

\paragraph{Funding information}
Supported by BMBF and the Helmholtz Association

\begin{appendix}



\end{appendix}





\begin{thebibliography}{99}
	
	
	\nolinenumbers
	
%
	
	\bibitem{cgcIntro}
	F. Gelis, E. Iancu, J. Jalilian-Marian, and R. Venugopalan, 
	\textit{The Color Glass Condensate}, 
	Ann. Rev. Nucl. Part. Sci. 60 (2010) 463–489, \doi{10.1146/annurev.nucl.010909.083629}
	
	\bibitem{fcelIntro}
	I. Vitev, 
	\textit{Non-Abelian energy loss in cold nuclear matter}, 
	Phys. Rev. C 75 (2007) 064906, \doi{10.1103/PhysRevC.75.064906}
	
	\bibitem{upgrades}
	D. Blau, 
	\textit{Performance of the ALICE electromagnetic calorimeters in LHC Runs 1 and 2 and upgrade projects}, 
	\url{http://arxiv.org/abs/2001.02928}
	
	\bibitem{alice}
	ALICE Collaboration, K. Aamodt et al., 
	\textit{The ALICE experiment at the CERN LHC},
	JINST 3 (2008) S08002, \doi{10.1088/1748-0221/3/08/S08002}
	
	\bibitem{blue}
	A. Valassi and R. Chierici,
	\textit{Information and treatment of unknown correlations in the combination of measurements using the BLUE method},
	Eur. Phys. J. C 74 (2014) 2717, \doi{10.1140/epjc/s10052-014-2717-6}
	
	\bibitem{tcm}
	A. Bylinkin, N. S. Chernyavskaya, and A. A. Rostovtsev,
	\textit{Predictions on the transverse momentum spectra for charged particle production at LHC-energies from a two component model}, 
	Eur. Phys. J. C75 (2015) 166, \doi{10.1140/epjc/s10052-015-3392-y}
	
	\bibitem{pPb8TeV}
	ALICE Collaboration, S. Acharya et al.,
	\textit{Nuclear modification factor of light neutral-meson spectra up to high transverse momentum in p-Pb collisions at $\sqrt{s_{\text{NN}}}$ = 8.16 TeV},
	\url{http://arxiv.org/abs/2104.03116}
	
	\bibitem{nCTEQ}
	K. Kovarik et al.,
	\textit{nCTEQ15 - Global analysis of nuclear parton distributions with uncertainties in	the CTEQ framework},
	Phys. Rev. D 93 no. 8, (2016), \doi{10.1103/PhysRevD.93.085037}
	
	\bibitem{dss}
	D. de Florian, R. Sassot, M. Epele, R. J. Hernández-Pinto, and M. Stratmann,
	\textit{Parton-to-pion fragmentation reloaded},
	Phys. Rev. D 91 no. 1, (2015) 014035, \doi{10.1103/PhysRevD.91.014035}
	
	\bibitem{aess}
	C. A. Aidala, F. Ellinghaus, R. Sassot, J. P. Seele, and M. Stratmann,
	\textit{Global analysis of fragmentation functions for $\eta$ mesons},
	Phys. Rev. D 83 (2011) 034002, \doi{10.1103/PhysRevD.83.034002}
	
	
	\bibitem{epps}
	K. J. Eskola, P. Paakkinen, H. Paukkunen, and C. A. Salgado,
	\textit{EPPS16: Nuclear parton	distributions with LHC data},
	Eur. Phys. J. C 77 no. 3, (2017), \doi{10.1140/epjc/s10052-017-4725-9}
	
	
	\bibitem{cgc}
	T. Lappi and H. Mäntysaari, 
	\textit{Single inclusive particle production at high energy from HERA data to proton-nucleus collisions},
	Phys. Rev. D 88, 114020 (2013), \doi{10.1103/PhysRevD.88.114020}
	
	
	\bibitem{fcel}
	F. Arleo, F. Cougoulic, and S. Peigné,
	\textit{Fully coherent energy loss effects on light hadron production in pA collisions},
	JHEP 09 (2020) 190, \doi{10.1007/JHEP09(2020)190}
	
	
	\bibitem{alice5TeVhadrons}
	ALICE Collaboration, S. Acharya et al.,
	\textit{Transverse momentum spectra and nuclear modification factors of charged particles in pp, p–Pb and Pb–Pb collisions at the LHC}.
	JHEP 11 (2018) 013, \doi{10.1016/j.physletb.2018.10.052}
	
	\bibitem{cms}
	CMS Collaboration, V. Khachatryan et al.,
	\textit{Charged-particle nuclear modification factors inPb–Pb and p–Pb collisions at $\sqrt{s_{\text{NN}}}$ = 5.02 TeV},
	JHEP 04 (2017) 039, \doi{10.1016/j.nuclphysbps.2017.05.098}
	
	\bibitem{alice5TeV}
	ALICE Collaboration, S. Acharya et al.,
	\textit{Neutral pion and $\eta$ meson production in p–Pb collisions at $\sqrt{s_{\text{NN}}}$ = 5.02 TeV},
	Eur. Phys. J. C 78 no. 8, (2018) 624, \doi{10.1140/epjc/s10052-018-6013-8}
	

	
\end{thebibliography}

\end{document}